\documentclass[prl,twocolumn,noshowpacs,noshowkeys,superscriptaddress]{revtex4}
\usepackage{rotating}
\usepackage{graphicx}
\usepackage{latexsym}
\usepackage{amsfonts,amsthm,bm}

\newcommand{\rmd}{\mbox{\rm d}}

\begin{document}

\title{The Universal Amplitude Ratio $\Gamma^{-}/\Gamma^{+}$ for Two-Dimensional Percolation}
\author{Iwan Jensen} 
\affiliation{ARC Centre of Excellence for Mathematics and Statistics of Complex Systems,
Department of Mathematics and Statistics, The University of Melbourne, Victoria 3010, Australia}
\author{Robert M. Ziff}
\affiliation{Michigan Center for Theoretical Physics and Department of Chemical Engineering,
University of Michigan, Ann Arbor, MI 48109-2136, USA}

\begin{abstract}
The amplitude ratio of the susceptibility (or second size-moment) for two-dimensional 
percolation is calculated by two series methods and also by
Monte-Carlo simulation. The first series method is a new approach 
based upon integrating approximations to the scaling function. The second 
series method 
directly uses low- and high-density series expansions of the susceptibility, going
to unprecedented orders for both bond and site percolation on the square lattice.
Putting all methods together we find a consistent value 
$\Gamma^{-}/\Gamma^{+} = 162.5 \pm 2$, 
a significant improvement over 
previous results that placed the value of this ratio variously in the range of 14 to 220.
\end{abstract}



\maketitle

Percolation is one of the fundamental problems in statistical mechanics
\cite{Essam80a,StaufferPercBook} and is perhaps the simplest system 
exhibiting critical behavior.
Through its mapping to the 
$q$-state Potts model (for $q \to 1$)
many theoretical predictions follow, such as exact critical 
exponents in two dimensions.  Yet 
many unanswered questions remain.
One of these is the value of amplitude ratios,
which represent universal ratios of quantities
related to integrals of the scaling function.
A great amount of work 
has been done investigating amplitude ratios of various systems, both to demonstrate 
that systems expected to be in a given universality class have the same ratios, and to 
determine their values accurately \cite{PHA91}.   The study
of amplitude ratios  remains an active area of 
research (e.g., \cite{CGR04,Delfino04,DG04,EG03,Seaton02,FMS01,SV01,Lee96}).

In this paper we study specifically the universal amplitude ratio $\Gamma^{-}/\Gamma^{+}$ for percolation, where $\Gamma^{-} $ and  
$\Gamma^{+}$ are the amplitudes of the second size-moment (also called the susceptibility)
in the low- and high-density phases, 
respectively. This ratio has been especially difficult to estimate accurately and the 
status of the estimates is very controversial. In \cite{PHA91} values
are quoted in a wide range from 14 to 220 based on numerical estimates from
Monte-Carlo
simulations and series expansions for various percolation models. Furthermore,
there are no reliable field-theoretical estimate for this quantity. 
Several years ago Delfino
and Cardy \cite{DC98} studied the $q$-state Potts model using methods from 
quantum field theory and predicted a value of 74.2
for percolation,
by extrapolating the results for $q = 2$, 3, and 4 to $q = 1$.
This was consistent with the numerical work
of Corsten et al. \cite{CJJ89}, who gave the value 75 $(+40,-26)$, 
but inconsistent with many other measurements \cite{PHA91}.
In this work,
we study bond and site percolation on the square lattice and use 
extensive exact enumerations to obtain estimates for $\Gamma^{-}/\Gamma^{+}$,
using two
different approaches: one a novel approach based upon
directly integrating approximations to the scaling function, and the
second a more conventional analysis of the high- and low-density
series for the susceptibility. The estimates for the amplitude ratio are consistent with the value  
$\Gamma^{-}/\Gamma^{+}=162\pm3$.
We also carried out a Monte-Carlo calculation, which gave
an almost identical value of $163 \pm 2$.
These results are a significant improvement on previous published numerical
estimates.  

Percolation models are commonly formulated in a lattice setting with the edges 
and/or vertices occupied (or vacant) with probability $p$ ( or $1-p$). In this paper we 
limit our study to bond and site percolation on the square lattice ${\mathbb Z}^2$. 
We shall refer to {\em occupied} edges and vertices as bonds and sites, respectively.  
Nearest neighbor bonds (sites) are said to be connected and clusters are 
sets of connected bonds (sites). The behavior of the model is controlled by the 
occupation probability (or density of bonds/sites) $p$. When $p$ is smaller 
than a critical value $p_c$ all clusters remain finite. Above $p_c$
there is a non-zero probability of finding an infinite cluster.
The critical occupation probability is know exactly for bond percolation,
$p_c=1/2$ \cite{Kesten80}, and to a high degree of numerical accuracy for site 
percolation, $p_c=0.59274621(13)$ \cite{NZ00}.
The average cluster size $S(p) \sim \Gamma^{-} (p_c-p)^{-\gamma}$, which diverges 
as $p \rightarrow p_c^-$, plays a role similar to a susceptibility, with
$\gamma=43/18$.

Percolation problems are closely related to the combinatorial problem of the 
enumeration of lattice animals, which are connected subgraphs of a lattice. 
The size of a lattice animal is the number of connected sites (or bonds). 
A vertex (or edge) is said to be a {\em perimeter site} if it is 
a nearest neighbor of a site in the lattice animal.
Series expansions for various percolation properties, such as  
the average cluster size, can be obtained as weighted sums over 
the number of lattice animals, $g_{s,t}$, enumerated according to the 
number of sites (bonds) $s$ and perimeter $t$. 
{\em Perimeter polynomials} are defined as 
\begin{equation}\label{eq:perimpol}
D_s(q) = \sum_t g_{s,t}q^t.
\end{equation}
We have calculated the 
perimeter polynomials up to size 35 (bond) and 40 (site).
The central property describing the cluster statistics in percolation is $n_s$, 
defined as the number of clusters (per site) containing $s$ occupied sites or bonds, 
as a function of the occupation probability $p$,
\begin{equation}\label{eq:clusnum}
n_s(p) = p^sD_s(1-p) = \sum_t g_{s,t}p^s (1-p)^t.
\end{equation}

The scaling hypothesis \cite{StaufferPercBook} states that
$n_s$ behaves as 
\begin{equation}\label{eq:ns}
n_s(p) = c_0s^{-\tau}f(c_1(p-p_c)s^{\sigma})  \;\;\;\;\; (p\to p_c, \; s\to \infty),
\end{equation}
where the critical exponents $\tau$ and $\sigma$ and the scaling function $f$ are 
universal, while $c_0$ and $c_1$ are non-universal metric factors.
In two dimensions $\tau = 187/91$ and $\sigma = 36/91$.
We shall often use the scaling variable $z=c_1(p-p_c)s^{\sigma}$. 
It then follows that \cite{StaufferPercBook}
\begin{eqnarray} \label{eq:Sint}
S(p) &=& \sum_{s} s^2n_s(p) 
      \propto  \int  c_0s^{2-\tau}f(c_1(p-p_c)s^{\sigma})\rmd s \nonumber \\
%      &=& \frac{c_0c_1^{(\tau-3)/\sigma}}{\sigma}|p-p_c|^{(\tau-3)/\sigma} 
 %         \int |z|^{-1+(3-\tau)/\sigma}f(z)\rmd z  \nonumber \\
      &=& \frac{c_0c_1^{-\gamma}}{\sigma}|p-p_c|^{-\gamma} 
          \int |z|^{-1+\gamma}f(z)\rmd z  \nonumber \\
      &=& \Gamma^{\pm} |p-p_c|^{-\gamma},
\end{eqnarray}
where $\gamma=(3-\tau)/\sigma$. 
The integration in the high- or low-density case extends from $0$ to $\infty$  or 
from $-\infty$ to $0$, respectively.  The amplitudes $\Gamma^{+}$ and $\Gamma^{-}$ are given
by the non-universal  constant $c_0c_1^{-\gamma}/\sigma$ multiplied by 
the corresponding integrals of
the universal scaling function $f(z)$. It thus follows that the ratio  $\Gamma^{-}/\Gamma^{+}$
is universal. 
Now, from the knowledge of $n_s(p)$ for finite $s$,
where $n_s(p)$ is a polynomial in $p$,
we can estimate the integrals in (\ref{eq:Sint}) by  approximating $f(z)$
by  $\bar{n}_s(z)= s^\tau n_s(z)$,
where $z=(p-p_c)s^{\sigma}$, and taking the ratio of the relevant integrals
involving $\bar{n}_s(z)$.  
Obviously, $\bar{n}_s(z)$ is just a polynomial approximation to $f(z)$ so we can't 
extend the integration to infinity (the integral would just diverge). However,
there is a natural cut-off provided by the scaling variable $z$ and the fact
that the physical low-density region is $0\leq p < p_c$, so the integral involving $z$
over the low-density region runs over the interval $[-z_-=-s^{\sigma}p_c,0]$.
Likewise, integrals over the high-density region include the interval 
$[0,s^{\sigma}(1-p_c)=z_+]$. The integrals are easily evaluated since 
$\bar{n}_s(z)=\sum_k a_kz^k$ are polynomials, which 
we can determine by exact enumeration.  We get in the high-density case:
\begin{eqnarray*}
I_s^+&=&\int_0^{z_+}\!\! |z|^{\gamma-1}\bar{n}_s(z)\rmd z =
  \int_0^{z_+} \!\!\!\!\! z^{\gamma-1} \sum_{k=0} a_kz^{k} \rmd z\\
 &=& \sum_{k=0} a_k (z_+)^{k+\gamma}/(k+\gamma),
\end{eqnarray*}
and in the low-density region
\begin{eqnarray*}
I_s^-&=&\int_{-z_-}^0 \!\!\!\!\!\!\! (-z)^{\gamma-1} \bar{n}_s(z)\rmd z =
  \int_0^{z_-}\!\!\!\!\!z^{\gamma-1} \sum_{k=0} a_k (-z)^{k} \rmd z \\ 
  &=& \sum_{k=0} (-1)^{k}a_k (z_-)^{k+\gamma}/(k+\gamma),
\end{eqnarray*}
from which we obtain the estimate $\Gamma^{-}_s/\Gamma^{+}_s=I_s^-/I_s^+$. We
shall call this ratio $r_s$ for short. The amplitude ratio $\Gamma^{-}/\Gamma^{+}$
is obtained from the limit of $r_s$ as $s\to \infty$.

\begin{figure*}[t]
   \centering
   \includegraphics[scale=0.65]{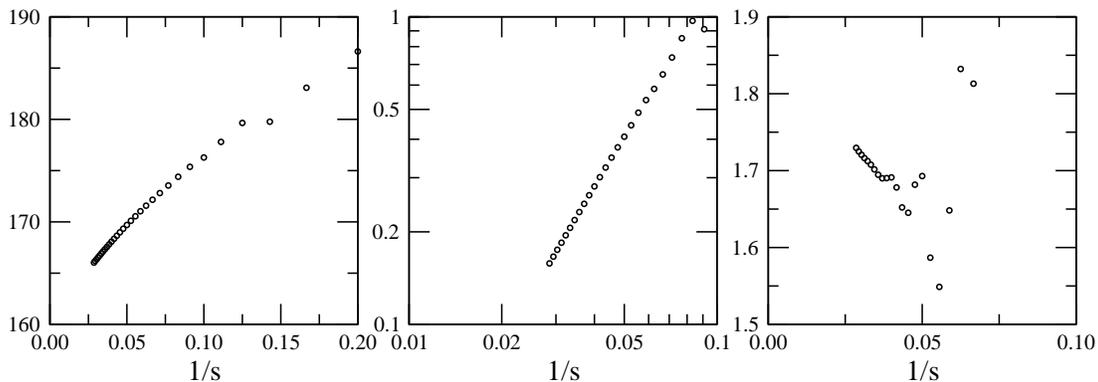} 
   \caption{The amplitude ratio  $r_s$ for bond percolation (left panel),
    a log-log plot of the difference between consecutive ratios (middle panel), and
    the local exponent of the difference plot (right panel).}
   \label{fig:amplratbond}
\end{figure*}

In Fig.~\ref{fig:amplratbond} we show (in the left panel) the estimates of
the ratio $r_s=I_s^-/I_s^+$ for bond percolation. The estimates display some
curvature when plotted against $1/s$, but seem to become close to a straight line
and extrapolate to a value around 160. In order to extrapolate to the limit $s\to \infty$ 
with a bit more confidence we need to try and work out the asymptotic form of
$r_s$. In the middle panel we show a log-log plot of the difference between consecutive 
ratios, $d_s=r_s-r_{s-1}$, against $1/s$. Clearly, $d_s$ has a power-law decay with $1/s$. 
This means that $r_s \approx \Gamma^{-}/\Gamma^{+} + a/s^{\alpha}$, and is is straightforward 
to show that then $d_s\propto 1/s^{\alpha+1}$. We now try to estimate the decay exponent 
$\alpha$. In the right panel of  Fig.~\ref{fig:amplratbond} we plot estimates for the 
{\em local exponent} $\alpha_s$ against $1/s$, where $\alpha_s$ is obtained from a linear 
regression of $\log d_s$ vs.~$\log s$ using the 5 values from $s$ to $s-4$. From this figure 
we estimate that $\alpha+1=1.85(5)$.

\begin{figure}
   \centering
   \includegraphics[scale=0.33]{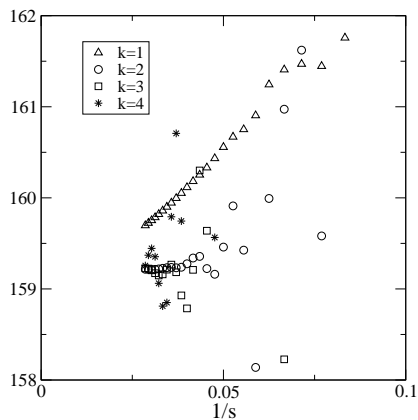} 
   \caption{Estimates of the amplitude ratio $\Gamma^{-}/\Gamma^{+}$ for bond percolation 
    from fits to the form (\protect{\ref{eq:amplfit}}) with $\alpha=0.85$.}
   \label{fig:amplextbond}
\end{figure}

Next we extrapolate the data for $r_s$ by fitting
directly to an assumed asymptotic form
\begin{equation}\label{eq:amplasymp}
r_s= \frac{\Gamma^{-}}{\Gamma^{+}} +\sum_{i\geq 0} \frac{a_i}{s^{\alpha_i}},
\end{equation}
where the exponents $\alpha_i$ form a strictly increasing sequence. By way of justification 
we can mention that similar forms are commonly found in the study of the asymptotic behavior of
series coefficients and arise from corrections to scaling. We don't know
the values of $\alpha_i$, except for the leading exponent $\alpha_0=\alpha$, so we
 assume that $\alpha_i=\alpha+i$ (this is akin to including
only one non-analytic correction to scaling), and fit to the form
\begin{equation}\label{eq:amplfit}
r_s=\frac{\Gamma^{-}}{\Gamma^{+}} +\sum_{i=0}^{k-1} \frac{a_i}{s^{\alpha+i}}.
\end{equation}
That is, we take a sub-sequence of terms $\{r_s,r_{s-1},\ldots,r_{s-k}\}$, plug into the formula 
above and solve the $k+1$ linear equations to obtain estimates for the amplitudes.
It is then advantageous to plot estimates for the leading amplitude $\Gamma^{-}/\Gamma^{+}$ against 
$1/s$ for several values of $k$. The results using the value $\alpha=0.85$ are plotted in 
Fig.~\ref{fig:amplextbond}. We notice that the estimates obtained with
$k=2$, 3, and 4 are quite stable and show little dependence on $s$ for large values.
This would indicate that $r_s$ is well approximated by the assumed asymptotic form.
We estimate from this that  $\Gamma^{-}/\Gamma^{+} = 159.2 \pm 0.2$,
where the error bars represent fluctuations but not systematic error
associated with this method. We also tested the sensitivity of 
the extrapolation procedure to the value of $\alpha$ by fitting to the form (\ref{eq:amplfit})
using the values $\alpha=0.8$ and $0.9$, respectively. The resulting estimates
for the amplitude ratio were only marginally lower or higher  and the procedure is not
very sensitive to changes in $\alpha$, at least within a reasonable range.

A similar analysis was carried out for site percolation. In this case the estimates 
for $r_s$ displayed a pronounced curvature when plotted against $1/s$ and the extrapolation
to the limit $s\to \infty$ was more challenging than in the bond case. We estimate that 
 $\Gamma^{-}/\Gamma^{+} = 164.5 \pm 1.5$. 

For our second approach, we have obtained estimates for the amplitudes $\Gamma^{-}$ and $\Gamma^{+}$ 
directly from the low- and high-density series for the second size-moment $M_2(p)$. In 
the low-density region the series was simply obtained from the perimeter polynomials 
$$M_2(p) = \sum_s s^2p^s \sum_t (1-p)^t g_{s,t} \sim \Gamma^{-}(p_c-p)^{-\gamma},$$
while the high-density series was calculated separately to order 51 (bond) and 55 (site). 
Estimates for the amplitudes were obtained by transforming the original series,
using our knowledge of the critical point and exponents, into series which have
a simple pole at $p_c$, with a residue from which we can calculate the amplitude.
The following two methods are quite standard \cite{AJG89a}. We can raise the series to the power
$1/\gamma$ to get a series with the behavior $(\Gamma^{-})^{1/\gamma}/(p_c-p)$ or
we can look at the series $M_2(p)(p_c-p)^{\gamma-1}$ which should have the behavior
$\Gamma^{-}/(p_c-p)$. The amplitudes can then be estimated by forming Pad\'e approximants
to the transformed series and calculating the residues. 
%In all the gory details we
%proceeded as follows. We start with a transformed series, say $M_2(p)^{1/\gamma}$,
%and we then form a Pad\'e approximant to this series. That is we find polynomials
%$P_m(p)$ and $Q_n(p)$ of degree $m$ and $n$, respectively, such that the series
%for the ration $P_m/Q_n$ agrees with the transformed series to order $m+n+1$ (we
%can assume that $Q_n(0)=1$). We now find the zero $p'$ of $Q_n$ closest to $p_c$
%and form the new polynomial $\tilde{Q}_n(p)$ defined by $Q_n(p)=(p'-p)\tilde{Q}_n(p)$.
%The amplitude of the transformed series is estimated as $P_m(p')/\tilde{Q}_n(p')$ (notice
%that we could have replaced $p'$ with $p_c$ in this calculation). 
Another approach is to get completely rid of the physical singularity at $p_c$ by studying
the transformed series $M_2(p)(p_c-p)^{\gamma}$ or $M_2(p)^{1/\gamma}(p_c-p)$ and evaluating
Pad\'e approximants to this series at $p_c$. As pointed out by Daboul {\em et al.} in their
recent study of 2D percolation \cite{DAS00}, such biased methods for calculating the amplitudes
generally lead to more accurate estimates when judged purely on the spread among the
estimates as obtained from individual approximants. However, this higher apparent
accuracy can be misleading in that biased approximations can display systematic
corrections on a scale exceeding the fluctuations. Thus great care must be taken 
and one should be careful not to rely too heavily on the biased estimates, particularly 
when it comes to determining the error bars on the estimates.

\begin{figure}
   \centering
   \includegraphics[scale=0.7]{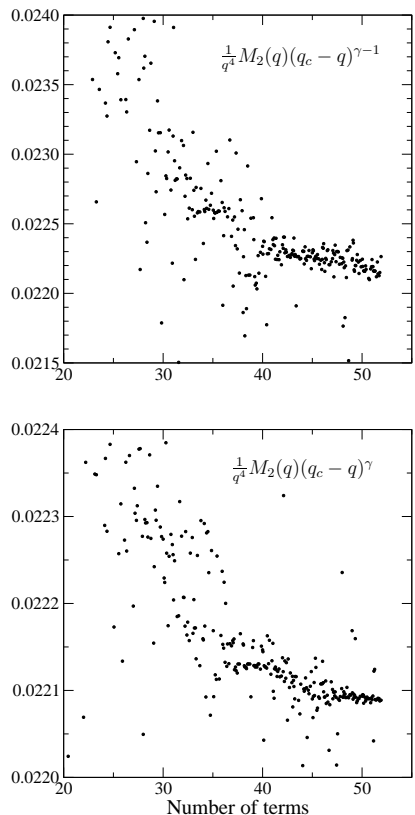} 
   \caption{The high-density amplitude $\tilde{\Gamma}$ for site percolation.}
   \label{fig:sqssdampl}
\end{figure}

In Fig.~\ref{fig:sqssdampl} we have plotted estimates for the high-density amplitude 
$\tilde{\Gamma}$ for site percolation. The estimates are obtained from Pad\'e approximants 
to the series $\frac{1}{q^4}M_2(q)(q_c-q)^{\gamma-1}$ (upper panel) and  
$\frac{1}{q^4}M_2(q)(q_c-q)^{\gamma}$ (lower panel), where the factor $1/q^4$ is included
in order that the transformed series has a non-zero constant term. 
As expected
the biased estimates (lower panel) are quite well-converged and appear to settle
down at a value around $0.02208(2)$. This is obviously slightly at odds with
the unbiased estimates (upper panel) which favor a higher value, but these estimates have 
a greater spread and a pronounced downwards drift. Estimates from the other methods
display similar trends and in particular have a greater spread than those in the
lower panel. To be cautious we adopt the rather conservative estimate
$\tilde{\Gamma}=0.02215(15)$ and from this we get $\Gamma^{+} = q_c^4 \tilde{\Gamma}=0.000609(4)$.
Similarly we estimate that in the low-density case
$\Gamma^{-} = 0.09819(6)$ and thus $\Gamma^{-}/\Gamma^{+} = 161.2 \pm 1.2$.
From a similar analysis we estimate that the amplitudes of the second 
size-moments of bond percolation in the low- and high-density 
phases are $\Gamma^{-} =0.15001(8) $ and 
$\Gamma^{+} = 0.0009290(15)$, which yields the
estimate  $\Gamma^{-}/\Gamma^{+} = 161.5 \pm 0.4$ for the amplitude ratio.

The estimates from the series studies are consistent with the value $\Gamma^{-}/\Gamma^{+}=162\pm3$, 
which is a significant improvement over previous results. 
We have also carried out a 
Monte-Carlo test of this amplitude ratio, studying bond percolation on the
square lattice. 
The method used was to generate individual clusters by a growth
algorithm on a large lattice at $p = 0.47$, 0.48, 0.49, $0.495$, $0.505$,
0.51, 0.53, and 0.53.  For $p < p_c$,
all the clusters terminate, while for $p > p_c$, the clusters that keep
growing beyond a clear cutoff can be identified as being part of the
infinite cluster and discarded.  This method proved superior to previous MC works
which generally count clusters on fully
populated lattices and have significant finite-size effects; here,
there were essentially no finite-size effects as long as $p$ is
kept sufficiently away from $p_c$ and the lattice is made large
enough.  This work yielded the result
$\Gamma^-/\Gamma^+ = 163 \pm 2$, which was quoted previously
in \cite{DBC00}.  

The two determinations of the amplitude ratio (by series and Monte-Carlo)
were done separately and the analysis of each was 
done by the two authors independently.  Putting these two unbiased
results together, we propose a final value
of $162.5 \pm 2$, where
the estimated error is at the 68\% confidence interval. 
We have thus found consistent precise series-expansion
and Monte-Carlo
results, laying to rest the controversy on the value of this amplitude
ratio.  Most likely, the wide spread in previous measurements 
were due to the large finite-size effects in Monte-Carlo
simulations, and the relatively short series used in exact
enumerations.  We have eliminated both of those shortcomings
in the present work.

As part of a larger on-going project we have also 
calculated perimeter polynomials on the honeycomb lattice, and a preliminary analysis
yields the estimates $\Gamma^{-}/\Gamma^{+}=166\pm 5$ (bond) and $170 \pm 7$ (site).
Finally, we mention in passing that we have studied the {\em area} moments
of bond percolation on the square lattice (where the area is the volume
enclosed by the external hull walk) (see \cite{CZ03} for details) and find
that the amplitude ratio of the second area-moment  is $175 \pm 10$. This value
is close to the one for the second size-moments, but it is expected
to have a somewhat different value because it represents a different
moment of the scaling function   $f(z)$.

This research has been supported by the Australian Research Council (IJ) and
the NSF (RMZ) under the grant DMS-0553487.
The calculations presented in this paper used the computational resources of the
Australian Partnership for Advanced Computing (APAC) and the  
Victorian Partnership for Advanced Computing (VPAC).

%\bibliographystyle{physrev}
%\bibliography{animals,percolation,sap,series}

\end{document}